\documentclass[aps,10pt, twocolumn, pra, superscriptaddress, nolongbibliography]{revtex4-2}
\usepackage{bm}
\usepackage{amsmath}
\usepackage{graphicx}
\usepackage{amsfonts}
\usepackage{siunitx}

\usepackage[normalem]{ulem}

\usepackage[caption=false,position=b,singlelinecheck=off]{subfig}

\def\ket#1{\mathinner{|{#1}\rangle}}

\def\Bra#1{\mathinner{\left\langle{#1}\right|}}
\def\Ket#1{\mathinner{\left|{#1}\right\rangle}}

\usepackage{color}

\begin{document}

\title{Improving trapped-ion qubit memories via code-mediated error-channel balancing}
\author{Yannick Seis}
\affiliation{Niels Bohr Institute, Blegdamsvej 17, 2100 Copenhagen \O, Denmark}
\affiliation{Center for Hybrid Quantum Networks (Hy-Q), Niels Bohr Institute, Blegdamsvej 17, 2100 Copenhagen \O, Denmark}

\author{Benjamin J. Brown}
\affiliation{Niels Bohr Institute, Blegdamsvej 17, 2100 Copenhagen \O, Denmark}
\affiliation{Centre for Engineered Quantum Systems, University of Sydney, Sydney, New South Wales 2006, Australia}

\author{Anders S. S\o rensen}
\affiliation{Niels Bohr Institute, Blegdamsvej 17, 2100 Copenhagen \O, Denmark}
\affiliation{Center for Hybrid Quantum Networks (Hy-Q), Niels Bohr Institute, Blegdamsvej 17, 2100 Copenhagen \O, Denmark}

\author{Joseph F. Goodwin}
\affiliation{Clarendon Laboratory, Parks Road, Oxford, OX1 3PU, United Kingdom}

\date{\today}

\begin{abstract}
  The high-fidelity storage of quantum information is crucial for quantum computation and communication. Many experimental platforms for these applications exhibit highly biased noise, with good resilience to spin depolarisation undermined by high dephasing rates.
  In this work, we demonstrate that the memory performance of a noise-biased trapped-ion qubit memory can be greatly improved by incorporating error correction of dephasing errors through teleportation of the information between two repetition codes written on a pair of qubit registers in the same trap.  While the technical requirements of error correction are often considerable, we show that our protocol can be achieved with a single global entangling phase gate of remarkably low fidelity, leveraging the fact that the gate errors are also dominated by dephasing-type processes.
  By rebalancing the logical spin-flip and dephasing error rates, we show that for realistic parameters our memory can exhibit error rates up to two orders of magnitude lower than the unprotected physical qubits, thus providing a useful means of improving memory performance in trapped ion systems where field-insensitive qubits are not available.
\end{abstract}

\maketitle

\section{Introduction}

The realisation of quantum technologies will require long-term maintenance of coherent quantum states. Both noisy-intermediate-scale quantum algorithms and future fault-tolerant codes are most often evaluated with respect to their performance in the presence of gate errors alone. However, in large-scale devices certain qubits will remain idle for extended periods, due to limits on the parallelism of gate operations across the qubit register~\cite{Das21} - both intrinsic to the circuit or code, or due to constraints of the architecture.
In this context, memory performance becomes a crucial consideration beyond its impact on gate fidelity, and the associated storage errors can become a significant barrier to the performance of the wider system~\cite{Chen21, Jurcevic21, Das21}. Beyond computational applications, such performance is also of crucial concern in repeater nodes, where coherence must be preserved for extended periods while entanglement is established between the end-point nodes~\cite{Azuma22, Kimble2008}. In addition to developing error correction codes for the protection against operational errors it is thus highly desirable to also develop codes tailored for long term storage of information.

Memory decoherence is typically characterised by the lifetime or depolarisation time of the logical states $T_1$ and their coherence time $T_2$. While coherence is fundamentally limited by depolarisation ($T_2\leq2T_1$), in most qubit systems dephasing rates far exceed this minimum, biasing the memory  to some greater or lesser degree towards Pauli $Z$-type errors~\cite{Debroy21, Robertson17}. In this manuscript, we will consider trapped ion qubits encoded in hyperfine- or Zeeman-split ground state manifolds. These systems typically exhibit collision-limited $T_1$ times of many thousands of seconds, while $T_2$ times due to magnetic field noise are far lower, typically 1-100 ms~\cite{sepiol19, goodwin16, Tan15, ruster16}, corresponding to a factor $10^4$--$10^7$ bias between phase-flip ($Z$) and bit-flip ($X/Y$) error channels. While the mechanisms responsible for the bias vary, similar imbalances are commonplace across a wide range of qubit platforms, e.g. dissipatively-stabilised superconducting cat qubits can exhibit biases of up to $10^9$~\cite{Lescanne2020, berdou22}.

Considerable work has been dedicated to improving the coherence times of individual ions by minimising the effects of magnetic field noise using, for instance, field-independent qubits~\cite{Langer05, sepiol19}, shielding~\cite{ruster16}, and dynamical decoupling~\cite{wang17}. These interventions can be highly effective, but are not universally applicable, nor without cost. The use of clock qubits is restricted to certain isotopes, many less accessible (e.g. $^{133}\textrm{Ba}^+$~\cite{Hucul17}) or less easily manipulated (e.g. $^{43}\textrm{Ca}^+$~\cite{Weber22, Allcock16}), and the use of these qubits is associated with increased gate and mapping errors. Magnetic shielding can greatly enhance the performance of field-sensitive qubits, but a considerable noise bias remains, and other sources of dephasing are unaffected. Dynamical decoupling methods can yield significant improvement, but are only effective in suppressing correlated noise and have no impact on memoryless or stochastic processes.

In this Manuscript we propose a simple protocol to extend the coherence time of an encoded qubit, utilising repetition codes written on a chain of trapped ions. By teleporting the logical information periodically between two such codes and determining syndrome information in the process, we can detect dephasing errors before they grow large enough to introduce logical errors. It has long been recognised that strongly polarised noise can be highly desirable when devising schemes for error correction~\cite{Darmawan2021, Robertson17, Webster15}, and that repetition codes provide a particularly straightforward solution in this scenario. However, the significant experimental overheads associated with encoding, decoding and syndrome readout via two-qubit gates, combined with the fact that such codes degrade performance on the unprotected channels, have limited the interest in these simple constructions.

In our protocol, the experimental overheads are minimised via the use of single, global entangling operations, both to teleport the initial qubit state into one code, and to subsequently teleport the logical information repetitively between the two subsystems in the trap (each subsystem capable of supporting a repetition code). Syndrome information is efficiently detected in a single projective measurement of the qubits in the teleporting code. The use of a native interaction which generates the complete circuit necessary for logical teleportation greatly simplifies the protocol, reduces the associated error rate, and removes any need for local logical operations. Crucially, it also makes the process much faster, enabling repeated teleportation at a frequency sufficient to significantly reduce logical errors, even in the presence of high dephasing rates.

Crucial to the success of the protocol is the availability of a global entangling gate (as opposed to only nearest-neighbour two-qubit entangling gates) that preserves the noise bias of the qubits, i.e. a gate with very low Pauli-$X$ type errors. The $ZZ$ geometric phase gates that can be engineered in trapped ion systems provide an ideal candidate, and for appropriate choice of gate parameters the bit-flip error can be reduced to the $10^{-5}$-level. Due to the global nature of the gate, small imperfections in its execution will introduce an error on a given ion with a probability that scales with the size of the system, and indeed for larger codes the gate fidelity rapidly approaches zero. However, the Pauli-$Z$ type errors that dominate the gate performance are identified and corrected by the code we prepare and therefore do not present a significant problem. Instead, small imperfections in the entangling operation control parameters only marginally reduce the threshold of dephasing noise the code will tolerate, typically with negligible impact on performance.

The result of implementing this protocol is a logical qubit memory with balanced rates of error in each channel. For the realistic parameters we consider, this reduces the net logical error rate by up to two orders of magnitude, depending on the size of the codes used and the degree of bias in the entangling gate. Loosely analagous to a refresh cycle in classical computer memory, we believe this simple protocol can provide a vital boost to memory performance for idle qubits in larger processors, networks or repeater nodes.

This Manuscript is structured as follows. In Sec.~\ref{sec:setup} we describe the experimental setup, the operations we have available, and the states we use in our code. In Sec.~\ref{sec:protocol}, we give the explicit program to execute error correction.
In Sec.~\ref{sec:noise}, we provide some estimates of the projected performance of the scheme by means of a noise analysis.
Finally we give some concluding remarks and propose extensions in Sec.~\ref{sec:conclusion}. Appendices~\ref{app:a},~\ref{app:e},~\ref{app:b} and~\ref{app:d} give technical details of important calculations.

\section{Experimental Setup}
\label{sec:setup}

We consider a register of $N$ ions in a linear radiofrequency ion trap. The trap frequencies are generally adjusted such that the laser-cooled ions crystallise into a one-dimensional Coulomb crystal. We assume the trap is segmented allowing for the definition of multiple trapping zones arranged along the zero-point of the oscillating radio frequency (RF) trapping field, enabling splitting, merging and shuttling of ion crystals between zones. We will make use of these features to divide the register into two equally sized smaller partitions which will host the encoded information. These partitions can be physically separated via shuttling operations enabling addressing of each partition individually, or combined within a single trapping zone for operations addressing the entire register simultaneously.

We encode qubits in pairs of Zeeman or hyperfine sublevels in the $S_{1/2}$ ground states of each ion, with the particular sublevels depending on the ion species. We will typically write qubits using the eigenbasis of the Pauli-$Z$ operator, such that a qubit register is denoted $\ket{\mathbf{s}} = \ket{s_1 s_2 \dots s_N}$ where $s_j = 0,1$. We denote the standard Pauli matrices acting on the $j$-th qubit of the register as $X_j$, $Y_j$ and $Z_j$. States $\ket{\pm_x} = (\ket{0} \pm \ket{1})/ \sqrt{2}$ and $\ket{\pm_y} = (\ket{0}\pm \text{i} \ket{1})/\sqrt{2}$ denote the eigenstates of the $X$ and $Y$ matrices with eigenvalue $\pm1$, respectively.

State preparation of the qubits is achieved by optically pumping the ions on a strong dipole transition, driving the population into $\ket{0}$ or $\ket{1}$ as required. 
Projective measurements in the Pauli-$Z$ basis are performed by driving the qubit on the dipole transition and observing state dependent fluorescence. Collective local rotations $R_x(\theta)$ and $R_y(\theta)$ of all the qubits in the trap can be achieved via the application of radio-frequency (RF) or microwave radiation at the qubit frequency. By combining projective measurements and state preparations in the $Z$ basis with collective RF or microwave rotations of the whole register, it is possible to measure and prepare partitions in other bases as required.

Many-ion entanglement can be achieved via a generalised geometric phase gate, namely a periodic, state-dependent optical dipole force (ODF) driving the collective motional modes of the crystal. This ODF is typically produced via a pair of crossed, far-detuned Raman lasers, producing a one-dimensional optical lattice which scans across the crystal at the difference frequency of the two beams. If this frequency is slightly detuned from a vibrational mode resonance of the Coulomb crystal~\cite{wineland98, ozeri11}, the interaction entangles spin states to states of motion, executing state-dependent excursions in phase space, which periodically return to zero. At this point the spin states are left disentangled from the motion, but acquire a geometric phase related to the degree of excitation~\cite{leibfried03, ozeri11}.

By tuning the ODF drive frequency close to different vibrational modes of the crystal it is possible to achieve qubit couplings proportional to the amplitude of the motion of each ion in the collective eigenmodes. Applied to two-ion crystals, this qubit-phonon coupling provides the basis for most forms of bipartite entangling gate~\cite{Bruzewicz2019}. In previous work, some of us proposed the use of inhomogeneous but structured global couplings to many-ion crystals to produce complex entanglement in Penning traps, where the target states shared the symmetry of the driven mode~\cite{goodwin15}. The protocol described in this proposal also utilises global entangling operations, but uses excitations of the CoM mode alone. When driven by a force perpendicular to the crystal axis this couples homogeneously to all ions, allowing the approach to be applied to the linear chains common to most radiofrequency ion traps. As we only require control of the individual partitions during state preparation or readout, no local coherent control is required, avoiding the experimental complexity associated with high-fidelity addressing of single ions within a many-ion crystal.

\subsection{Unitary operations in ion traps}
\label{subsec:opps}

To generate the entanglement necessary for our protocol we require the ability to perform a global unitary operation $U$ such that, up to local $Z$ rotations on the individual spins, we have
\begin{equation}
U \ket{\mathbf{s}} = (-1)^{|\mathbf{s}|(|\mathbf{s}| - 1) / 2}  \ket{\mathbf{s}}, \label{eq:global}
\end{equation}
on basis states $\ket{\mathbf{s}}$ where $|\mathbf{s}|$ is the Hamming weight of $\mathbf{s}$, i.e., the number of spins of $\ket{\mathbf{s}}$ in the $\ket{1}$ state. In Appendix~\ref{app:a} we show that driving the CoM interaction for time $T = 2 \pi k / \delta$, with integer $k$, detuning $\delta = \mu_L - \omega_1$, Raman laser difference frequency $\mu_L$ (i.e. the periodic driving force frequency), CoM mode frequency $\omega_1$ and appropriately chosen ODF magnitude, allows us to implement this unitary operation. Specifically, we show that this allows us to implement $U'(T) \sim U $ (the two unitaries being equal up to single qubit operations) where
\begin{equation}
U'(T) = e^{-\text{i} H_{\text{int} } T }, \label{Eqn:UnitaryEvolution}
\end{equation}
with the interaction Hamiltonian for the CoM vibrational mode
\begin{eqnarray}
H_{\text{int}} = \frac{J}{N} \sum_{j < k}^N Z_j Z_k,  \label{eq:Hi:init}
\end{eqnarray}
and the coupling strength is given by
\begin{eqnarray}
J \approx \frac{F^2}{4 M \omega_1 \delta}, \label{eq:jjk:CoM}  
\end{eqnarray}
where each ion has mass $M$, and the travelling one-dimensional optical lattice imparts a periodic force on each ion with magnitude $F$. We also note that provided we keep track of the local $Z$ rotations that may occur we need not correct the single-qubit rotations. Indeed, as we will see, they do not alter the capability of the code to protect against dephasing errors.

\begin{figure}
\centering
\includegraphics[trim = 50 70 80 60 , clip, width=\columnwidth]{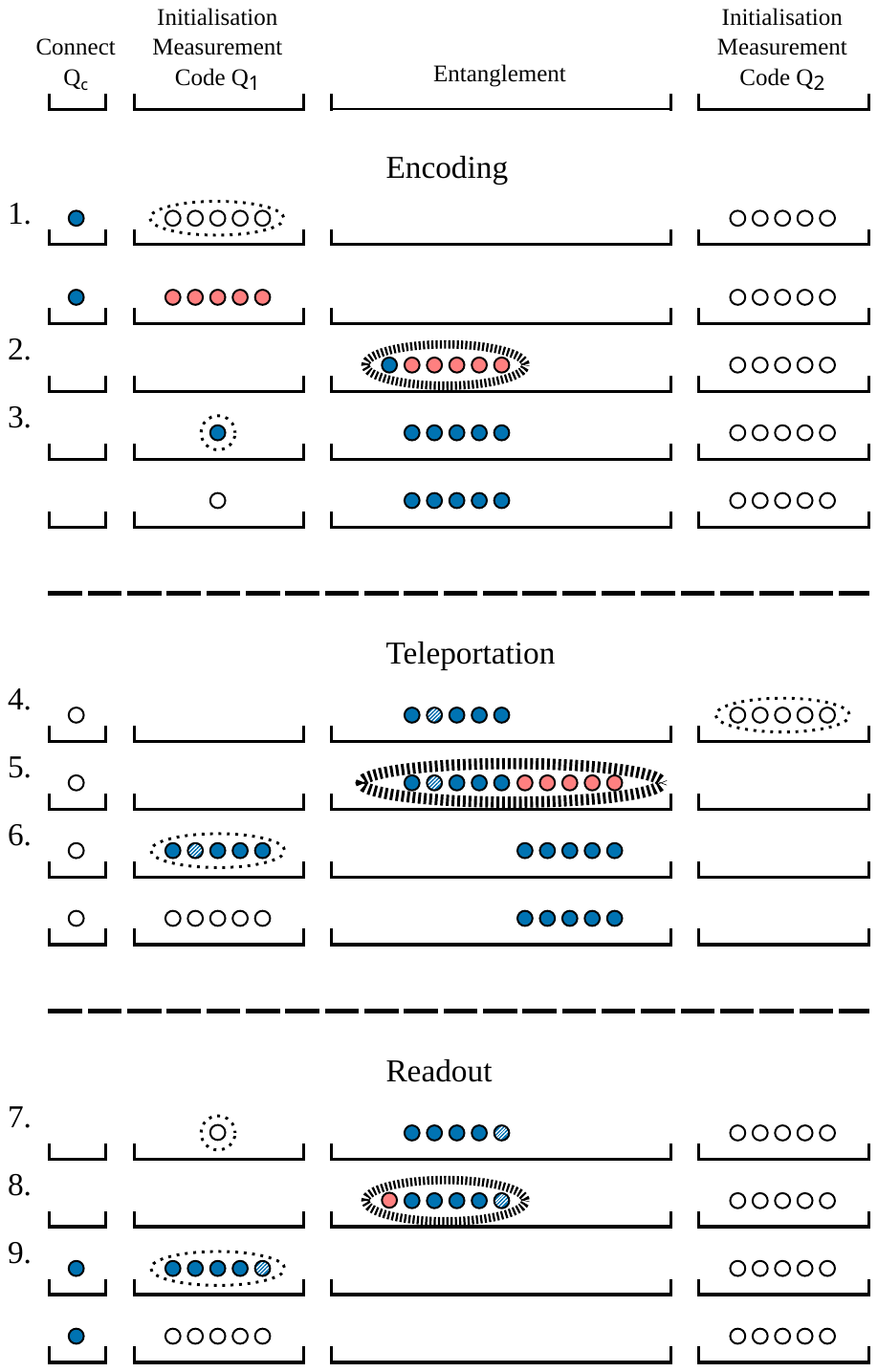}
\caption{
The experimental protocol performed in a linear radiofrequency ion trap with functionalised areas for qubit manipulation. The numbering corresponds to the protocol steps in Sec.~\ref{sec:protocol} of the main text. Qubits hosting the logical information are coloured in (dark) blue, with single qubit Z errors marked in striped (dark) blue, initialised qubits in (light) red and other qubits in the $\ket{0}$ state in white. In the ``Initialisation'' and  ``Measurement'' areas, thin dotted ellipses denote initialisation into the $\ket{+_x}$ states and measurement in the Pauli-X basis respectively. Thick dashed ellipses in the ``Entanglement'' area denote the global unitary $U$. \emph{Encoding:} The logical information is teleported from the (left-most) communication qubit $\mathcal{Q}_\text{c}$ to a repetition code on $\mathcal{Q}_1$, where it is better protected against $Z$ errors. \emph{Teleportation:} If $Z$ errors occur on $\mathcal{Q}_1$, we can teleport the information to a clean repetition code on $\mathcal{Q}_2$ and preserve the quantum state, provided a minority of physical qubits experienced this type of error. This step is repeated, moving the information between $\mathcal{Q}_1$ and $\mathcal{Q}_2$ alternately. \emph{Readout:} The quantum state can be read out by teleporting to the communication qubit. Again, small numbers of $Z$ errors on $\mathcal{Q}_1$ can be corrected for and are not tranferred to the communication qubit.}
\label{fig:protocol}
\end{figure}

\section{Protocol}
\label{sec:protocol}

Having introduced the experimental setup we now give an explicit program to carry out our protocol based on the repetition code. For a detailed discussion of the code and the teleportation of information between codes see Appendix~\ref{app:e}.

In the protocol, we partition the register of $N$ ions into two codes of size $n = N/2$. We teleport the information between two codes, one hosted on each of the partitions $\mathcal{Q}_1$ and $\mathcal{Q}_2$. We obtain syndrome data at each cycle from the collective local projective measurement of the qubits in the teleporting partition, enabling us to correct for $Z$ errors that might have occured on that partition post-initialisation, while keeping the logical information continuously protected with an error-correcting code.
The smallest implementation of the protocol can be performed in a linear crystal with $N \geq 6$ physical qubits (with codes of larger size giving better protection against dephasing errors), where each code partition hosts the smallest possible repetition code with $n=3$ qubits.

\subsubsection*{The protocol program}

We first define the code partitioning. The crystal is divided into three subsets $\mathcal{Q}_1, \mathcal{Q}_2$ and $\mathcal{Q}_\text{c}$ as is illustrated in Fig.~\ref{fig:protocol}. Partitions $\mathcal{Q}_1$ and $\mathcal{Q}_2$ support codes $1$ and $2$, respectively, while $\mathcal{Q}_\text{c}$ is a single communication qubit used to initialise the code. As discussed in Sec.~\ref{sec:setup}, by shuttling the qubits of each partition into separate trap zones, we can choose to perform state preparation and projective measurement on a single partion, while global rotations affect qubits in all zones equally.

In what follows we describe the steps to encode and read out logical information from the qubit register, together with the steps we follow to learn syndrome data while information is encoded on the register. Here we focus only on the procedure to perform the protocol; details explaining how the protocol works are given in Appendices ~\ref{app:a},~\ref{app:e} and~\ref{app:b}. Technical details on how the necessary physical operations can be implemented are discussed in Appendix~\ref{app:d}.

The following three steps first encode an arbitrary state onto a repetition code across the qubits of subset $\mathcal{Q}_1$ (see ``Encoding'' in Fig.~\ref{fig:protocol}, where step numberings coincide).

\begin{enumerate}
\item \label{list:Initialise} Initialise the crystal such that the qubits of $\mathcal{Q}_1$ are in $\ket{+_x}$, and the communication qubit $\mathcal{Q}_\text{c}$ in the state we wish to encode, $\ket{\psi_y}~=~\alpha\ket{+_y}+\beta\ket{-_y}$. 
\item Apply the global unitary $U$.\label{list:Couple}
\item Measure the communication qubit in the $Y$ basis to teleport the information onto $\mathcal{Q}_1$, up to a correction determined by the measurement outcome. \label{list:meas}
\end{enumerate}

By completing the first three steps we teleport the logical data onto a repetition code encoded on $\mathcal{Q}_1$. From here we can measure syndromes while maintaining the encoding by teleporting the information onto a second partition $\mathcal{Q}_2$, via entanglement of the two partitions and measurement of the qubits of $\mathcal{Q}_1$. Alternatively we can read out the information by teleporting the logical state back to the communication qubit, which provides us with syndrome data while decoding the information.
To read out the encoded information, we proceed to step~\ref{list:readout} below. To move the encoded information to a new partition, we continue with steps~\ref{list:firstSyndromeStep} to~\ref{list:lastSyndromeStep} (corresponding to ``Teleporation'' in Fig.~\ref{fig:protocol}). 

\begin{enumerate}
\setcounter{enumi}{3}
\item Prepare all of the qubits of $\mathcal{Q}_2$ in the state $\ket{+_y}$.  \label{list:firstSyndromeStep} 
\item Apply the $U$ operation. \label{list:USyndrome}
\item Measure all of the qubits of $\mathcal{Q}_1$ in the Pauli-X basis to teleport the state onto $\mathcal{Q}_2$ up to a correction determined by the majority outcome of the $X$ measurements. \label{list:lastSyndromeStep}
\end{enumerate}

The program between step~\ref{list:firstSyndromeStep} and step~\ref{list:lastSyndromeStep} teleports the protected information from $\mathcal{Q}_1$, which may have suffered phase errors, to the freshly-initialised state in $\mathcal{Q}_2$ which has not yet undergone decoherence. Any ions that have experienced dephasing errors will give a inverted measurement outcome in step~\ref{list:lastSyndromeStep}, providing the necessary syndrome. By determining the teleportation correction via a majority vote we effectively utilise this syndrome data, ensuring that the protocol preserves the logical state provided fewer than half the qubits experience errors. Similarly, if the information is stored on $\mathcal{Q}_2$, we can detect and correct for errors by performing the reciprocal operation where we teleport the information from $\mathcal{Q}_2$ to $\mathcal{Q}_1$.

Note that while we carry out steps~\ref{list:firstSyndromeStep} through~\ref{list:lastSyndromeStep}, the communication qubit remains idle. During this period, we are free to include the communication qubit within one of the code subsets provided it is once again isolated before proceeding to step~\ref{list:readout}.

In the final steps we read out information from the code, where in the following steps we suppose that information is stored in code 1. However, the steps are easily adapted if information is stored in code 2.  The ``Readout'' steps in Fig.~\ref{fig:protocol} are as follows.

\begin{enumerate}
\setcounter{enumi}{6}
\item Initialise the communication qubit to $\ket{+_y}$. \label{list:readout}
\item Apply $U$.
\item Measure the qubits of $\mathcal{Q}_1$ in the $X$ basis to teleport logical information back to the communication qubit, determining the appropriate Pauli correction via a majority vote of the measurement outcomes.
\end{enumerate}
With the logical information moved back to the communication qubit, we are free to extract this quantum information via single-qubit operations.

We remark that the state we encode is vulnerable to errors affecting the communication qubits before it is encoded in step~\ref{list:meas}. This is common to all schemes for encoding arbitrary states for any choice of quantum error-correcting code. Alternatively, there are certain special states that we can prepare on the code partition which are never left unprotected. We can encode logical $\overline{X}$ eigenstates $\ket{\overline{\Lambda_+(\vartheta)}}$ or $\ket{\overline{\Lambda_-(\vartheta)}}$ (where the overbars indicate the \emph{logical} qubit space, the subscripts the logical eigenstates, and $\vartheta$ the equatorial encoding basis of the physical qubits) by initialising all the qubits of $\mathcal{Q}_1$ into the $\ket{+_\vartheta}$ or $\ket{-_\vartheta}$ state, respectively, neglecting the communication qubit and skipping steps~\ref{list:Couple} and~$\ref{list:meas}$ - these encoded states are simply product states. Observing that our teleportation protocol applies a logical Hadamard operation at each step, we can see that it is also possible to initialise logical states $\ket{\overline{\Lambda_{0,1}(\vartheta)}}=(\ket{\overline{\Lambda_+(\vartheta)}} \pm \ket{\overline{\Lambda_-(\vartheta)}})/\sqrt{2}$ on $\mathcal{Q}_1$ by initialising all of the qubits in $\mathcal{Q}_1$ and $\mathcal{Q}_2$ in the $\ket{+_\vartheta}$ state, applying $U$, and measuring the syndromes on $\mathcal{Q}_2$. The result is then equivalent to having teleported the logical state $\ket{\overline{\Lambda_+(\vartheta)}}$ state that would have been on $\mathcal{Q}_2$ onto $\mathcal{Q}_1$.

\section{Noise Analysis}
\label{sec:noise}

We now investigate how our proposed protocol performs in the presence of environmental noise and imperfect operations.

\subsection{Noise model}
Qubits encoded in ions are usually much more vunerable to dephasing errors than spin flips. Assuming that the qubits dephase with a rate $\gamma_Z=1/T_2^*$, and depolarise at a rate $\gamma_X=1/T_1$, we define the \emph{noise bias} of the system as $\eta = \gamma_Z / \gamma_X$. We will work in units where $\gamma_Z = 1$, scaling our dynamics to the characteristic decoherence rate of a single, unprotected qubit. The actual value of $\eta$ will vary greatly between experiments, but for qubits encoded in magnetically-sensitive electronic ground state sublevels it is typically $\eta\approx10^4-10^7$.

Given these parameters, the noise incident on each ion is described by the noise map $\mathcal{E} = \mathcal{E}_Z \circ \mathcal{E}_X$, where $\mathcal{E}_X$~($\mathcal{E}_Z$) describes spin-flip (dephasing) noise acting on each individual ion of the system. At time $t$ after initialisation of a clean code subset, we write these maps as
\begin{equation}
\mathcal{E}_\sigma(\rho) = (1 - p_\sigma(t)) \rho /2 +  p_\sigma(t) \sigma \rho \sigma /2 , \label{Eqn:NoiseMap}
\end{equation}
where $\sigma = X, Z$ are the standard Pauli matrices and 
\begin{equation}
p_X(t) = (1-e^{-t / \eta} )/ 2,
\end{equation}
and
\begin{equation}\label{Eqn:singlePhase}
 p_Z(t) = (1-e^{-t } )/ 2
\end{equation}
are the probabilities of an $X$ or $Z$ error having occurred by time $t$. It is worth noting that the noise channels $\mathcal{E}_Z  $ and $ \mathcal{E}_X$ commute. We are therefore able treat the two channels separately. To simplify our analysis, we also frequently make the assumption that $p_X(t)$ and $p_Z(t)$ are small.

\subsection{Figure of merit}
\label{sec:figmerit}

Because the repetition code only protects against errors on one channel, it does not have an associated \emph{threshold} physical error probability, below which logical errors can be strongly suppressed. Instead, we define our figure of merit as the total logical error rate $\Gamma$, which we wish to minimise. We will consider three error mechanisms (which we define explicitely in subsection \ref{sec:errorrates} below) that contribute to $\Gamma$: the logical $\overline{X}$ error rate $\Gamma_{\overline{X}}$ of the code, the logical $\overline{Z}$ error rate $\Gamma_{\overline{Z}}$ of the code and the average logical error rate $\Gamma_T$ due to teleportations between codes.

The purpose of our codes is to improve our resilience to dephasing at the cost of an increased vulnerability to spin-flip errors. With careful choice of code size, the total rate of \emph{logical} errors $\Gamma=\Gamma_{\overline{Z}} + \Gamma_{\overline{X}} + \Gamma_T$ on the encoded state can be greatly reduced below the total physical error rate. The optimum performance is achieved when the rate of logical errors for both channels are approximately equal, $\Gamma_{\overline{X}} + \Gamma_T \approx \Gamma_{\overline{Z}}$. As we will now show, the improvement that can be achieved increases with the degree of noise bias but will be limited by the quality of the gate operations available. Note that the structure of the repetition code means that physical $X$ errors lead to logical $\overline{X}$ errors, and physical $Z$ errors (to half or more of the qubits) lead to logical $\overline{Z}$.

We foresee that for varying code sizes, there exists a minimum in $\Gamma$ using the following intuition.
If the number $n$ of ion qubits in each of the code partitions $\mathcal{Q}_1$ and $\mathcal{Q}_2$ is low, the logical information is vulnerable to phase-flips ($\Gamma_{\overline{Z}}$ is large), while spin-flips degrade the stored state by only a small amount ($\Gamma_{\overline{X}}$ is small), and teleportation operations are possible with high fidelity ($\Gamma_T$ is small). By increasing $n$, we improve the protection that the code offers against dephasing errors, while logical errors due to spin-flips and teleportation errors become more appreciable. If $n$ becomes too large, the overall logical error rate becomes dominated by these channels and starts to increase with $n$.

\subsection{Error rates of the teleporation protocol}
\label{sec:errorrates}

\begin{figure*}
\includegraphics[width=0.9\textwidth]{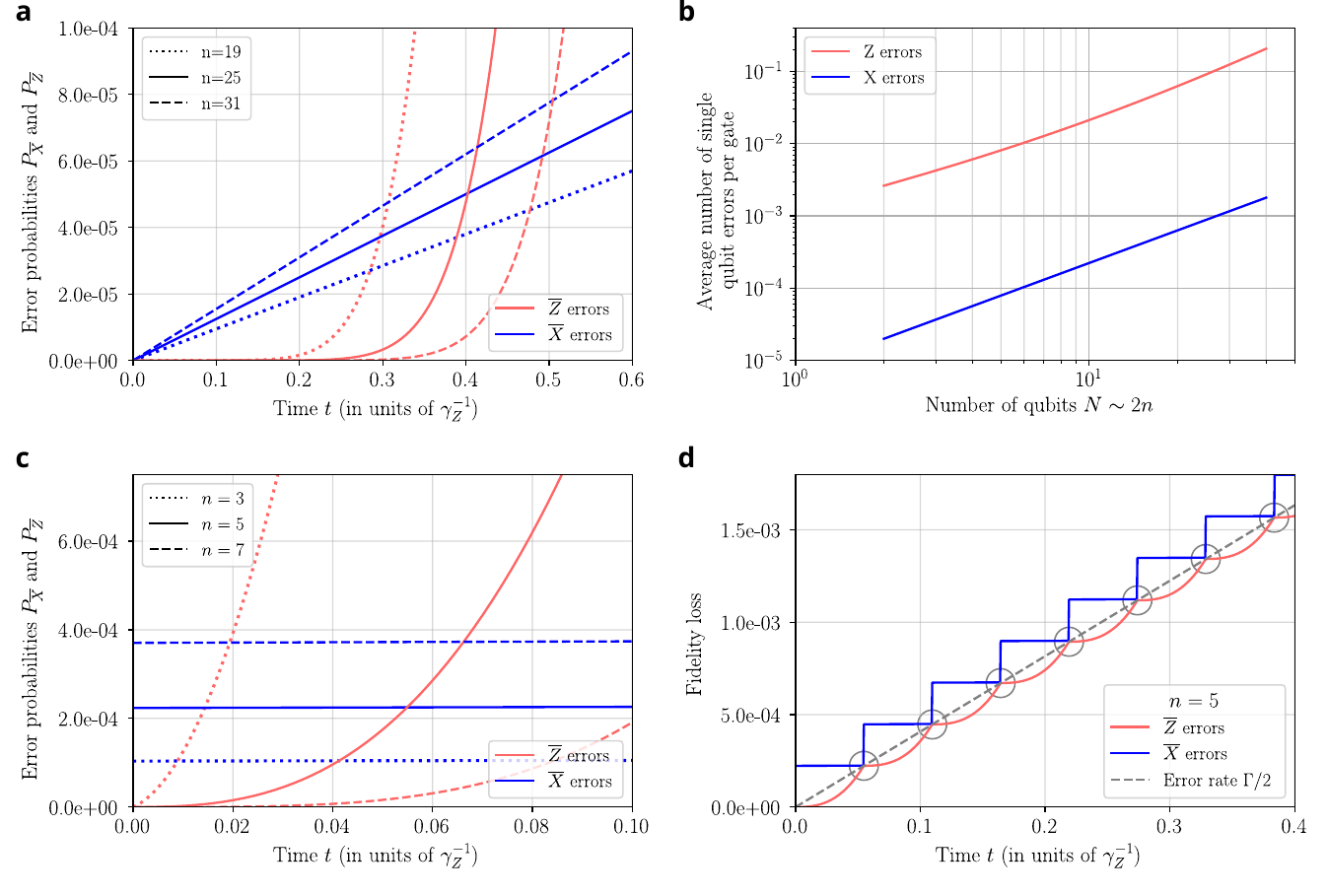}
\caption{Logical error probabilities.
   (a) Calculated logical $\overline{X}$ and $\overline{Z}$ error probabilities for codes of size $n=19$, $25$ and $31$ (assuming $\eta = 10^5$), when we do not consider any teleporation induced errors. The curves correspond to Eqs.~(\ref{Eqn:BitFlipRate}) and (\ref{Eqn:PhaseFlipRate}) and arrows indicate increasing $n$.
  (b) Average total number of physical $X$ and $Z$ errors occuring across the register during each teleportation process as function of the number of qubits being entangled, $N$. Gate errors are strongly biased toward Z, and both error channels scale as approximately $\mathcal{O}(N^{3/2})$ for large $N$, due to the $\sqrt{N}$ scaling of gate duration. The assumed error rate corresponds to a two-qubit gate fidelity of 99.7\% (see Appendix~\ref{app:d} for full details of error model). We emphasise that physical $X$ errors lead to logical $\overline{Z}$ errors and vice-versa, with the colour-coding consistent across the subfigures.
  (c) Calculated logical $\overline{X}$ and $\overline{Z}$ error probabilities for codes of size $n=3$, $5$ and $7$ ($\eta = 10^5$), where we now include the teleporation induced errors given in b). Teleportation errors manifest as a non-zero $\overline{Z}$ probability at $t = 0$. Note that compared to a), we must use much lower numbers of qubits to keep error probabilities low.
  (d) Fidelity loss as the information is repeatedly teleported between codes of size $n=5$ for $\eta=10^5$. The circles correspond to teleportation events, which are here chosen to be when $P_{\overline{X}}$ and $P_{\overline{Z}}$ are equal. The dashed line corresponds to an average error rate $(\Gamma/2)$ for each of the $\overline{X}$ and $\overline{Z}$ channels.
}
\label{Fig:Nepserr}
\end{figure*}

With the repetition code protecting against phase errors, the logical information is lost once a single spin flip is introduced to the system. 
The probability $P_{\overline{X}}(t,n)$ that a single spin has flipped scales linearly with $n$, as described by the first-order expression
\begin{equation}
P_{\overline{X}}(t,n)  \approx n p_X(t) \approx \frac{nt}{2 \eta}    \label{Eqn:BitFlipRate}
\end{equation}
assuming that each ion of the system is subject to the noise map $\mathcal{E}_X$ given in Eq.~(\ref{Eqn:NoiseMap}) and $p_X(t)$ is small.

We can correct dephasing errors introduced to the repetition code provided fewer than half of the qubits have experienced dephasing. The probability that a majority of qubits suffer a dephasing error is
\begin{align}
  P_{\overline{Z}}(t,n) = \sum_{j=n_e}^n \begin{pmatrix} n\\ j\end{pmatrix} p_z(t)^j(1-p_z(t))^{n-j}, \label{Eqn:PhaseFlipRate}
\end{align}
where each ion is subject to the noise map $\mathcal{E}_Z$ and $n_e = (n-1)/2$ when $n$ is odd and $n_e = n/2 - 1$ when $n$ is even.
For illustration, we plot $P_{\overline{X}}(t,n)$ and $P_{\overline{Z}}(t,n)$ for varying numbers of physical qubits in Fig.~\ref{Fig:Nepserr}a.

So far the probabilities $P_{\overline{X}}$ and $P_{\overline{Z}}$ only looked at the errors occuring while the information is already encoded in the entangled state. To these error probabilities we also add the errors induced by the entangling operation. In Fig.~\ref{Fig:Nepserr}b we plot the probabilities for $X$ and $Z$ errors as function of the number of the entangling qubits. Teleportation-induced phase-flip errors will reduce the number of qubits effectively available in the code~\footnote{In general, $n_\text{eff}$ is not an integer. In this case, we compute $P_{\overline{X}}(t, n_\text{eff})$ as a weighted sum of the probabilities for the integers closest to $n_\text{eff}$. For example, $P_{\overline{X}}(t, n=15) \rightarrow P_{\overline{X}}(t, n_\text{eff} = 14.9) = 0.1P_{\overline{X}}(t, n = 14) + 0.9 P_{\overline{X}}(t, n = 15)$.}: if on average, say, one qubit experiences a phase-flip during the teleportation this effectively reduces the code size from $n$ to $n_\text{eff} = n-1$, although we note that for the code sizes considered $n-n_\text{eff}\ll1$. We see here that while the teleporation operation induces some $Z$ errors in the physical qubits, our code is able to mitigate them.
On the other hand, bit-flip errors during the teleportation, which we denote $P_T(N)$, are not correctable and add to the total logical error rate as shown below. Note that in considering entangling gate error probability, we must consider all the $N$ qubits involved in the operation, whereas for memory or measurement errors we consider $n \sim N/2$ qubits.

In our protocol we propose to teleport the logical information many times between two codes. Given the probabilities of logical error per cycle, we can express the total logical error rate as
\begin{align}
\Gamma = \frac{P_{\overline{Z}}(T_\text{tele},n_\text{eff}) + P_{\overline{X}}(T_\text{tele},n) + P_T(2n)}{T_\text{tele}}, \label{eq:TotalErrRate}
\end{align}
where $T_\text{tele}$ is the time between two teleportation operations.

This expression for $\Gamma$ reveals a trade-off in logical error rates for increasing $T_\text{tele}$. For fixed $n$, the logical $\overline{X}$ error rate $\Gamma_{\overline{X}}~=~P_{\overline{X}}(T_\mathrm{tele},n) / T_\mathrm{tele}$ is constant in time and equal to $n/2\eta$, as seen from Eq.~(\ref{Eqn:BitFlipRate}). However the logical $\overline{Z}$ error rate $ \Gamma_{\overline{Z}}~=~P_{\overline{Z}}(T_\text{tele},n_\text{eff}) / T_\text{tele}$ grows with $T_\text{tele}$, while the teleportation error rate $\Gamma_T = \frac{P_T(2n)}{T_\text{tele}}$ decreases with $T_\text{tele}$.
For a register with $2n$ (or $2n+1$) qubits, the decoherence rate $\Gamma$ is thus minimised by teleporting the logical information between the two code subsets at certain period $T_\text{tele}$.

\begin{figure*}
\includegraphics[trim = 0 0 0 0 , clip, width=0.9\textwidth]{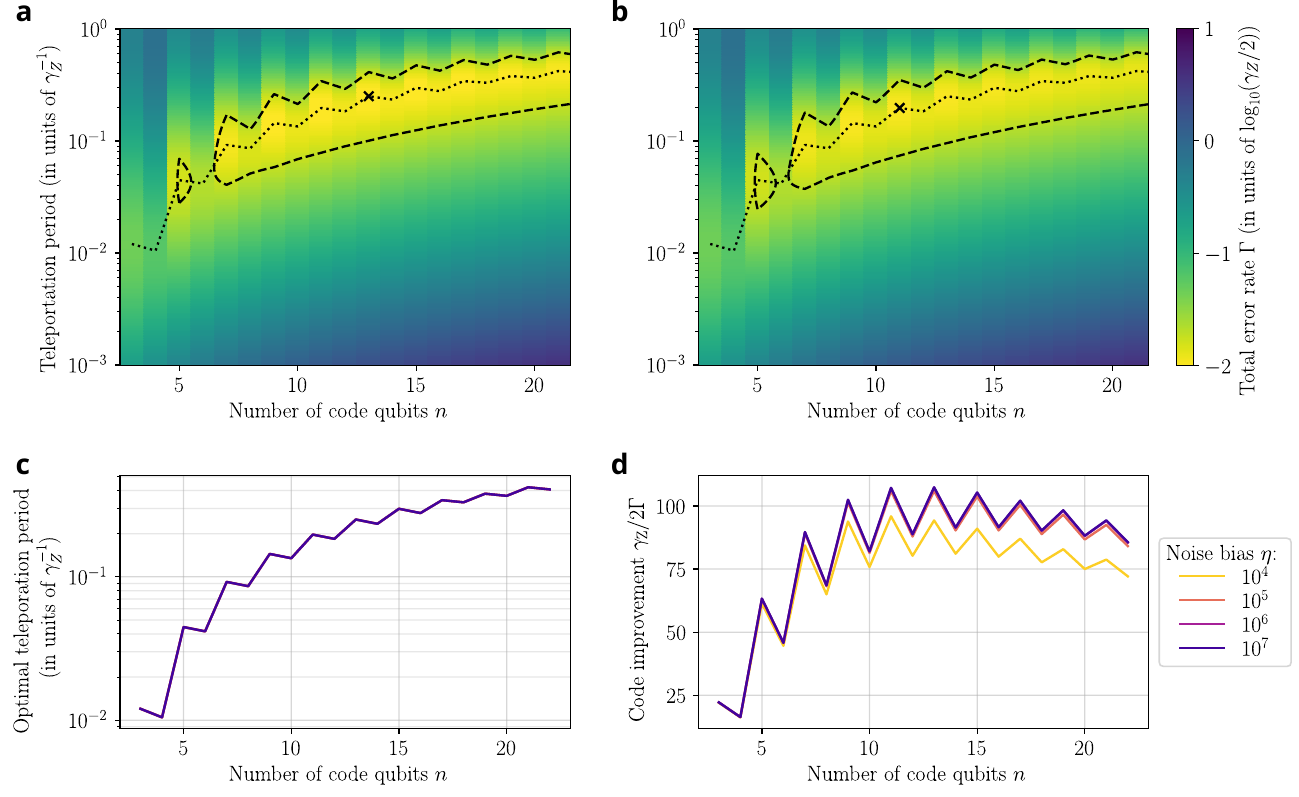}
\caption{Minimisation of the total error rate $\Gamma$ and protocol robustness. Using the teleportation error model plotted in Fig.~\ref{Fig:Nepserr}b, we plot the error rate when varying the number $n$ of qubits in the code subsets and the time $T_\mathrm{tele}$ (in units of $\gamma_\mathrm{Z}^{-1}$) between teleporation events for (a) $\eta = 10^6$ and (b) $\eta = 10^4$ (subplots (a) and (b) share the same colorbar).
  The black crosses indicate $n$ and $T_\mathrm{tele}$ where $\Gamma$ is minimised ($\Gamma = \Gamma_\mathrm{min}$), giving an improvement of almost two orders of magnitude versus the unprotected qubit error rate. The dashed contour line encloses the region $\Gamma\geq2\Gamma_\mathrm{min}$, emphasising that our protocol gives good improvements for a large range of code sizes and teleportation periods. The dotted line representes the choice of $T_\mathrm{tele}$ which minimises $\Gamma$ for a given $n$. 
  (c) The optimal value of the teleportation period $T_\mathrm{tele}$ in units of $\gamma_\mathrm{Z}^{-1}$ for a range of $n$ and $\eta$. Note that the curves for varying $\eta$ are almost perfectly identical, indicating that in the considered regime the protocol is limited by gate errors and independent of the spin flip rate $\gamma_x$. For typical trapped-ion bit-flip times of \SI{10000}{\s}, with an error bias $\eta=10^6$, i.e. a phase-flip time of \SI{10}{\ms}, the information should ideally be teleported every \SI{0.12}{\ms} (\SI{0.45}{\ms}) in codes of size $n=3$ ($n=5$).
  (d) The code improvement $ \gamma_\mathrm{Z} / 2 \Gamma$ yielded by codes of varying $n$ and $\eta$, when $T_\mathrm{tele}$ is chosen to minimise $\Gamma$. For the smallest implementations with codes of $n=3$ and $5$, the improvement factors are $22$ and $60$ respectively.
}
\label{Fig:map}
\end{figure*}

Note that for large noise bias, such as the example of $\eta=10^5$ given in Fig.~\ref{Fig:Nepserr}d, the system exhibits virtually no fidelity loss due to $\overline{X}$ errors. Instead, logical errors occur during teleportation and due to insufficient protection against physical dephasing, leading to large $\Gamma_T$ and $\Gamma_{\overline{Z}}$ respectively. The performance of the protocol is optimised when $\Gamma_T\approx\Gamma_{\overline{Z}}$, and improving the efficiency of the protocol mainly relies on the reduction of physical $X$ errors during the global entangling operation, allowing teleportation to be performed more frequently, and/or between higher-$n$ codes.

\subsection{Protocol robustness and performance}

Finally, let us compute the total error rate $\Gamma$ as function of the number of qubits in the code and of the teleportation period for low ($\eta=10^4$) and high ($\eta=10^6$) values of the noise bias.

In Fig.~\ref{Fig:map}a and b, we plot $\Gamma$ as function of the number $n$ of qubits in the code and the time $T_\mathrm{tele}$ between teleporations.
Firstly, for both values of $\eta$ there is a region of low error rate for suitable combinations of $n$ and $T_\mathrm{tele}$: as $n$ increases, the optimal $T_\mathrm{tele}$ must also go up to keep the contribution of $\Gamma_T$ to $\Gamma$ relatively low.
As already alluded to in Subsec.~\ref{sec:figmerit}, to the left of this high-performance region (low $n$), the stored information is vulnerable to $Z$ errors on the physical qubits.
Similarly from arguments in Subsec.~\ref{sec:errorrates}, below this region (i.e. for small $T_\mathrm{tele}$) $\Gamma_T$ dominates in $\Gamma$, whereas above it (large $T_\mathrm{tele}$) $\Gamma_{\overline{Z}}$ dominates the total error rate.

Secondly, as illustrated graphically in Fig.~\ref{Fig:Nepserr}d, the region of improved storage corresponds to a choice of teleportation period that ensures the overall logical error rates for the $\overline{X}$ channels ($\Gamma_{\overline{X}}, \Gamma_T$) and $\overline{Z}$ channel ($\Gamma_{\overline{Z}}$) are approximately balanced. Thirdly, Figs.~\ref{Fig:map}a and b show that we can achieve this large improvement in storage time over a wide range of $n$ and $T_\mathrm{tele}$ values, which makes the protocol robust to experimental constraints and imperfections.

In Fig.~\ref{Fig:map}c we plot the optimal teleporation time which minimises the total error (where curves for different $\eta$ overlap). For $\eta = 10^6$, the teleporation period ranges from \SI{0.1}{\ms} at low $n$ to \SI{\sim 3}{\ms} at $n=20$, when taking $\gamma_\mathrm{X}^{-1} = \SI{10000}{\s}$.

Finally, to assess the code performance, we compare the total error rate $\Gamma$ to the single qubit error rate $\gamma_\mathrm{Z}/2$. For a given code size $n$ and value of $\eta$, we compute the best code improvement $ \gamma_\mathrm{Z} / 2\Gamma$ in Fig.~\ref{Fig:map}d, where we take the lowest value of $\Gamma$ optimised over $T_\mathrm{tele}$. This optimal configuration is represented by a dashed line in Figs.~\ref{Fig:map}a and b.
The projected error rates achieved with this scheme are in the region of 10-100 times lower than the dephasing error rate $\gamma_Z/2$ of the individual physical qubits, far beyond the break-even point. Note that the largest relative improvements with increasing code size occur for the smallest implementations of the code ($n=3,5$), making significant gains in performance readily accessible to experimental realisation in current systems.

\section{Conclusion}
\label{sec:conclusion}

To summarise, we have proposed a simple yet efficient protocol to encode logical quantum information by continuously teleporting between two repetition codes on noise-biased trapped-ion qubits. The strength of the scheme lies in the use of a single global unitary operation for fast entanglement of the two code subspaces, and allows for simple syndrome readout by a majority vote on the measured qubits. Furthermore, we have made use of the fact that the dominant error type introduced by the teleporation operation is also the type corrected by the repetition encoding. The noise bias preserving nature of the teleportion complements the choice of physical qubits. For the considered high fidelity gates it can lead to an improvement in coherence time of up to two orders of magnitude by rebalancing the rates of the logical Pauli $X$ and $Z$ errors.

We envision the use of such a protocol as an intervention between the level of the physical qubits and higher-level quantum network or computing algorithms, to improve the memory performance of idling qubits, functionally analogous to the memory refresh cycle in classical dynamic random-access memory (DRAM). The results of our analysis emphasise that in architectures with heavily biased noise we can achieve significant memory improvements without the need for a full quantum error-correcting code and the overhead in resource and complexity that it entails.

For some applications, such as quantum network repeater nodes, this improvement in memory may be sufficient; in others it may be necessary to embed the stored information in a larger code. In both the establishment of long-distance entanglement across a multi-node network and in the generation of large codes via stabiliser measurements, it is necessary to produce and combine graph states via the application of $ZZ$ gates between qubits. Because the global operation we propose generates the logical $\overline{ZZ}$ natively, such graph states can be assembled from qubits stored in multiple repetition codes, leaving the information protected throughout. This is of particular utility when the graphs to be generated involve many bipartite links, and many qubits within the state must thus be held idle during assembly (e.g. for point-to-point communication via long chains of repeater nodes, or for high-weight stabiliser or resource-state assembly in distributed computational architectures).

Finally, we note that the modest number of physical qubits required ($N\geq6$) and the simple protocol make our proposal readily realisable in current trapped-ion experiments, with the potential for significant improvement in memory performance.

\begin{acknowledgments}
The authors would like to thank T. Rudolph for initially suggesting YS's Masters project, which evolved into the present manuscript. YS received support from ERC Grant Q-CEOM (638765) and changed affiliation to ENS de Lyon during the finalisation of this manuscript; BJB received support from the Villum Foundation, the University of Sydney Fellowship Programme and the Australian Research Council via the Centre of Excellence in Engineered Quantum Systems (EQuS) (CE170100009) and changed affiliation to IBM Quantum during the preparation of this manuscript; AS is supported by the Danish National Research Foundation (Hy-Q Centre of Excellence); JFG is supported by the United Kingdom Engineering and Physical Sciences Research Council “Networked Quantum Information Technology” (EP/M013243/1) and “Quantum Computing and Simulation” (EP/T001062/1) Hubs, and European Union Quantum Technology Flagship Project “AQTION” (820495). For the purpose of Open Access, the author has applied a CC BY public copyright licence to any Author Accepted Manuscript version arising from this submission.
\end{acknowledgments}

\appendix

\section{The centre-of-mass interaction}
\label{app:a}

In this Appendix we show that we can realise the unitary operator $U$ shown in Eq.~\eqref{eq:global} by applying the interaction Hamiltonian given in Eq.~(\ref{eq:Hi:init}) for a time $T = 2 \pi k / \delta$.

We consider a system with a spin-dependent ODF whereby ions in the $\ket{0}$ ($\ket{1}$) state experience a force $F_0$ ($F_1$). The motion of the crystal is excited, conditional on the internal spin states. If the excitation frequency is detuned from the CoM mode frequency $\omega_1$ by $\delta$, the crystal returns to its initial motional state after a time $t=2\pi/\delta$ (assuming $\delta\ll\omega_1$ so that the rotating wave approximation holds)~\cite{ozeri11}. This leads to the effective interaction Hamiltonian
\begin{equation}
H_\text{int} = \frac{J}{N} \sum_{j < k}^N Z'_j Z'_k , \label{eq:Hi:initAPP}
\end{equation}
where 
 \begin{equation}\label{eq:modZ}
 Z'  =\frac{1}{F} \begin{pmatrix} F_{0} & 0 \\ 0 & F_{1} \end{pmatrix},
\end{equation}
and $ J \approx F^2 / 4 M \omega_1 \delta $, with $N$ the number of ions in the crystal, $\omega_1$ the CoM mode frequency, $M$ the ion mass and $F = (F_0 - F_1)/2$ the strength of the differential force applied on the ion crystal by the optical travelling lattice. The entangling operation is least sensitive to laser intensity noise if we set $F_0=-F_1=F$, leading to the Hamiltonian defined in Eq.~\eqref{eq:Hi:init}. However other factors may constrain us to applying unbalanced forces. For this calculation it will be convenient to write $Z'_j$ in terms of Pauli matrices such that $Z'_j  =  R \openone_j +  Z_j $ where the parameter $ R = (F_0 + F_1) / (F_0 - F_1)$ captures the force imbalance.

To determine the entangled state which we produce we consider the energy eigenvalues of $H_{\text{int}}$. Given that $H_{\text{int}}$ is written in terms of Pauli-Z matrices, it is diagonal in the computational basis (with $\mathbf{s}$ labeling bit strings in the computational basis and $|\mathbf{s}|$ is the Hamming weight of the bit string). Using that $Z'_j \ket{s_j} = (R +1 - 2 s_j) \ket{s_j}$, we satisfy the energy eigenvalue expression $H_\text{int} \ket{\mathbf{s}} = \lambda_\mathbf{s} \ket{\mathbf{s}}$ with
\begin{equation}
\frac{ N \lambda_\mathbf{s}}{J  } =   4 \sum_{j<k}^N s_j s_k  -  A \sum_j^N s_j   + C ,    \label{eq:binand}
\end{equation}
where
\begin{equation}
A = 2  (R+1) (N-1)\;\;\text{and}\;\;C = \frac{N (N- 1)(R+1)^2}{2},
\end{equation}
yielding the unitary evolution $e^{-\text{i} H_\text{int} t } \ket{\mathbf{s}} = e^{-\text{i} \lambda_\mathbf{s} t } \ket{\mathbf{s}}  $.

To produce the desired unitary evolution in Eq.~(\ref{eq:global}), we must then verify that we can find system parameters such that
\begin{equation}
e^{-\text{i} \lambda_\mathbf{s} T } = V (-1)^{|\mathbf{s}| (|\mathbf{s}|-1) / 2} \label{eq:elambdas}
\end{equation}
for all values of $\mathbf{s}$ up to local rotations on the spins, denoted by $V$. To find parameters fulfilling Eq.~(\ref{eq:elambdas}), we use the identity
\begin{equation}
 \sum_{j<k}^N s_j s_k  = |\mathbf{s}| (|\mathbf{s}| - 1) / 2,
\end{equation}
and adjust the ODF to set $J$ such that $4J T / N  = \pi $~\footnote{In fact, we are free to take $4J tT  / N  = \pi (1 + 2l) $ for arbitrary values of $l$ where $l$ is an integer. Ideally, we will choose a value of $l$ that will minimise error.}. We then have
\begin{equation}
e^{-\text{i} H_\text{int} T } \ket{\mathbf{s}} = V U \ket{\mathbf{s}}, \label{Eqn:IntermediatePhase}
\end{equation}
where 
\begin{equation}
V =  \exp( ( - \text{i} ( - A |\mathbf{s}|   + C) \pi  / 4 ).
\end{equation} 
Ignoring the irrelevant global phase of $C\pi/4$, the additional unitary $V$ introduces a local $Z$ rotation on every physical qubit, which will not be problematic in our protocols provided we can keep accurate account of its magnitude. We discuss this in Apps.~\ref{subsection:codewords}~\&~\ref{subsection:NullingRotations}.

\section{The repetition code}
\label{app:e}

In this Appendix we give explicit details on the repetition code. We describe how to teleport encoded quantum states between two separate repetition codes to learn the locations of incident dephasing errors using the global operation and single-qubit measurements described in Sec.~\ref{subsec:opps}. We also show that we can encode arbitrary states to the repetition code using the same operations.

\subsection{Codewords}
\label{subsection:codewords}

We consider repetition codes of $n$ physical qubits that protect one logical qubit against dephasing errors. Encoded states, or codewords, are spanned by general equatorial basis states
\begin{equation}\label{Eqn:Logicalphi}
\begin{aligned}
\Ket{\overline{\Lambda_+(\vartheta)}} =  \underbrace{\ket{+_\vartheta} \ket{ +_\vartheta} \dots \ket{ +_\vartheta}} _n, \\
\Ket{\overline{\Lambda_-(\vartheta)}} = \underbrace{\ket{-_\vartheta } \ket{-_\vartheta } \dots \ket{ -_\vartheta}}_n,
\end{aligned}
\end{equation}
where $\ket{\pm_\vartheta}=(\ket{0}\pm e^{i \vartheta}\ket{1})/\sqrt{2}$ and $\vartheta$ takes an arbitrary value. We follow the convention that we denote vectors in the logical basis with a bar above the vector label. We thus write a logical qubit encoded in an arbitrary state as
\begin{equation} 
\Ket{\overline{\Lambda_\psi(\vartheta)}} = \alpha \Ket{\overline{\Lambda_+(\vartheta)}} + \beta \Ket{\overline{\Lambda_-(\vartheta)}}. \label{eq:repy}
\end{equation}

Note that we define $x = 0$ and $y = \pi / 2$ such that $\ket{\pm_x} \equiv\ket{\pm_{\vartheta = 0}}$ ($\ket{\pm_y} \equiv \ket{\pm_{\vartheta = \pi / 2}}$) are eigenstates of the Pauli-X(Pauli-Y) operators, as defined in the main text.

Dephasing errors affect physical qubits prepared in basis states $\ket{\pm_\vartheta}$ as follows
\begin{equation}
Z\ket{\pm_\vartheta } = \ket{\mp_\vartheta}.
\end{equation} 
As such encoded states will have equal protection against dephasing errors for any value of $\vartheta$. We maintain the freedom to vary the parameter $ \vartheta $ of encoded states as the $V$ term in Eq.~(\ref{Eqn:IntermediatePhase}) will change its value. The rotation of the basis is of no concern when applying our global single qubit rotations provided the value of $ \vartheta $ is known. The presence of $V$ is therefore only problematic if there is a relative uncertainty $\epsilon_{F}$ in our ODF strength $F$ which leads to an uncertainty in the phase of $\epsilon_\vartheta=(2\epsilon_{F}+\epsilon_{F}^2)$. In Appendix~\ref{app:d} we show how such errors can be nulled to first order through the use of appropriate ODF configurations.

\subsection{Syndrome readout by teleportation}

We first show that we can use a global entangling operation together with single-qubit measurements to learn syndrome information. We begin with two codes prepared in the initial state
\begin{eqnarray}\label{eq:init:b}
\ket{C} = \Ket{\overline{\Lambda_\psi(\vartheta)}}_1 \Ket{\overline{\Lambda_+(\varphi)}}_2. \label{Eqn:InitialStateBeforeTeleportation}
\end{eqnarray}
where the encoded information is stored on the first code, indexed $1$, and we prepare a second code, indexed $2$, in the logical state $\ket{\overline{\Lambda_+(\varphi)}}$ as in Eq.~(\ref{Eqn:Logicalphi}). Each code can take an arbitrary size and they do not necessarily have to use the same number of physical qubits. Note that this includes the special case where code 1 consists of a single qubit, allowing us to write and read arbitrary single qubit states to and from the code.

We can read syndrome data by entangling two codes in the trap and teleporting the information onto the second code with single-qubit measurements. We can entangle the codes in Eq.~(\ref{Eqn:InitialStateBeforeTeleportation}) with the global entangling operation which acts like a controlled-phase gate on the logical space of the two codes. One can check that
\begin{eqnarray} \label{Eqn:EntangledLogicalState}
&\overline{CZ} \ket{C} \approx \Ket{\overline{\Lambda_+(\vartheta)}}_1 \left( \frac{ \alpha+\beta}{\sqrt{2}} \Ket{\overline{\Lambda_+(\varphi)}}_2 + \frac{\alpha - \beta}{ \sqrt{2}} \Ket{\overline{\Lambda_-(\varphi)}}_2 \right)  \nonumber \\ 
&+\Ket{\overline{\Lambda_-(\vartheta)}}_1  \left( \frac{ \alpha+\beta}{\sqrt{2}} \Ket{\overline{\Lambda_+(\varphi)}}_2 - \frac{\alpha - \beta}{ \sqrt{2}} \Ket{\overline{\Lambda_-(\varphi)}}_2 \right).\;\;\;
\end{eqnarray}
where $\overline{CZ}$ is the controlled phase gate explicitly defined in Eq.~(\ref{Eqn:CZ}). We prove the equivalence between $U$ and $\overline{CZ}$ in Appendix~\ref{app:b}, up to local rotations that are not detrimental to our procedure for syndrome readout.

Now that the register is in the entangled state given in Eq.~\eqref{Eqn:EntangledLogicalState}, we measure the physical qubits in code $1$ in the $\ket{\pm_{\vartheta}}$ basis to teleport the logical information onto code $2$. If no dephasing errors have occured, the measurements will collapse all the qubits in code $1$ onto a common eigenstate, either $\ket{+_{\vartheta}}$ or $\ket{-_{\vartheta}}$, and thus either all ions will fluoresce or none of them will. The logical information will be successfully transferred onto the second subsystem up to a single-qubit unitary rotation, namely a Hadamard rotation on the logical qubit. This unitary squares to the identity operator and thus cancels after an even number of teleportations are performed.

Considering now the case where a few dephasing errors have occured, the affected qubits will collapse to the orthogonal state to that of the majority and exhibit correspondingly different fluorescence. If the majority of ions fluoresce, we assume the post-measurement state to be $\ket{\overline{\Lambda_+(\vartheta')}}_1$ where the prime indicates our inference that this corresponds to the initial state of code $1$ having been $\ket{\overline{\Lambda_+(\vartheta)}}_1$ and this having experienced a small number of dephasing errors. Similarly if the majority are dark, we assume the state to be $\ket{\overline{\Lambda_-(\vartheta')}}_1$, i.e. the state $\ket{\overline{\Lambda_-(\vartheta)}}_1$ after a small number of dephasing errors~\cite{goodwin15}. In this way, the teleporting measurement provides syndrome information telling us whether or not we need to apply a logical correction to recover the original encoded state. This syndrome information will only fail if more than $(n-1)/2$ of the qubits in code $1$ have suffered a dephasing error (in which case we would apply the wrong correction, introducing a logical error). Each time we perform a teleportation the logical data is mapped to recently initialised qubits, free of errors, so the probability of failure is kept low.

\section{Teleporting logical information between two repetition codes}
\label{app:b}

Here we show that the global operation $U$, as defined in the main text in Eq.~(\ref{eq:global}), will generate the entanglement we require to teleport the logical information from one repetition code to another through single-qubit measurements as we have already presented in Eq.~(\ref{Eqn:EntangledLogicalState}). We will show this statement holds for the case where the two repetition codes have different numbers of qubits, $n_1$ and $n_2$, respectively. It will be convenient to denote the two subsets of qubits of the system $\mathcal{Q}_\alpha$ with $\alpha = 1,2$. One repetition code is written to each of the two qubit subsets.

It is readily checked that Eq.~(\ref{Eqn:EntangledLogicalState}) holds provided we can show, up to local rotations on the logical space, that
\begin{equation}
U \Ket{\overline{\Lambda_A(\vartheta)}}_1 \Ket{\overline{\Lambda_B(\varphi)}}_2 \approx (-1)^{AB} \Ket{\overline{\Lambda_A(\vartheta')}}_1 \Ket{\overline{\Lambda_B(\varphi')}}_2, \label{Eqn:TargetExpression}
\end{equation}
where we rewrite encoded states in the basis
\begin{equation}
\Ket{\overline{\Lambda_A(\vartheta)}}_\alpha =  \frac{ \Ket{\overline{\Lambda_+(\vartheta)}}_\alpha + (-1)^A \Ket{\overline{\Lambda_-(\vartheta)}}_\alpha }{  \sqrt{2}} , \label{Eqn:SuperpositionCodeStates}
\end{equation}
where $A,\,B = 0,\,1$ denotes the state of the logical degree of freedom and $\alpha = 1,2$ specifies the subset of qubits $\mathcal{Q}_\alpha$ to which a code is written. The states $\ket{\overline{\Lambda_+(\vartheta)}}$ and $\ket{\overline{\Lambda_-(\vartheta)}}$ are defined in Eqs.~(\ref{Eqn:Logicalphi}).

To show that Eq.~\eqref{Eqn:TargetExpression} holds, it is convenient to express the basis states given in Eq.~\eqref{Eqn:SuperpositionCodeStates} in the computational basis. Up to normalisation coefficients we have
\begin{equation}
\Ket{\overline{\Lambda_0(\vartheta)}}_\alpha = \sum_{ \text{even }  |\mathbf{s}| } e^{\text{i} \vartheta |\mathbf{s}| } \ket{\mathbf{s}}_\alpha, \label{Eqn:CodeStates1}
\end{equation}
and
\begin{equation}
\Ket{\overline{\Lambda_1(\vartheta)}}_\alpha = \sum_{ \text{odd } |\mathbf{s}| } e^{\text{i} \vartheta |\mathbf{s}| } \ket{\mathbf{s}}_\alpha, \label{Eqn:CodeStates2}
\end{equation}
where the symbol $\mathbf{s}$ labels bit strings of length $n_\alpha$ and $|\mathbf{s}|$ is the Hamming weight of the bit string.

We will prove Eq.~(\ref{Eqn:TargetExpression}) by applying $U$ to pairs of the encoded basis states shown in Eqs.~(\ref{Eqn:CodeStates1}) and~(\ref{Eqn:CodeStates2}).
First we consider a simpler calculation where we find the action of $U$ when applied to unentangled states written in the computational basis. From Eq.~(\ref{eq:global}) we have
\begin{equation}
U\ket{\mathbf{s}}_1 \ket{\mathbf{t}}_2 = (-1)^{r(r-1)/2} \ket{\mathbf{s}}_1 \ket{\mathbf{t}}_2 \label{Eqn:IntermediateExpression}
\end{equation}
where $r = |\mathbf{s}| + |\mathbf{t}|$ and $\mathbf{s}$ and $\mathbf{t}$ are bit strings of length $n_1$ and $n_2$, respectively.

The result shown above in Eq.~\eqref{Eqn:IntermediateExpression} can be applied directly to the orthogonal states shown in Eqs.~\eqref{Eqn:CodeStates1} and~(\ref{Eqn:CodeStates2}) to prove Eq.~(\ref{Eqn:TargetExpression}). To do so, we expand the exponent of the eigenvalue found in Eq.~\eqref{Eqn:IntermediateExpression}. We find
\begin{equation}
\frac{r(r-1)}{2} = \frac{|\mathbf{s}|(|\mathbf{s}|-1)}{2} + \frac{|\mathbf{t}|(|\mathbf{t}|-1)}{2} +|\mathbf{s}||\mathbf{t}|, \label{Eqn:Expansion}
\end{equation}
where, importantly, the term $|\mathbf{s}||\mathbf{t}|$ takes an odd value if and only if both $ |\mathbf{s}|$ and $|\mathbf{t}|$ are odd.

Now, it follows from the discussion given above that
\begin{eqnarray}
U  \Ket{\overline{\Lambda_A (\vartheta)}}_1   \Ket{\overline{\Lambda_B(\varphi)}}_2  \nonumber & = & (-1)^{|\mathbf{s} || \mathbf{t}|}  \Ket{\overline{\Lambda_A'(\vartheta)}}_1 \Ket{\overline{\Lambda_B'(\varphi)}}_2 \nonumber \\ 
&=&   (-1)^{AB}  \Ket{\overline{\Lambda_A'(\vartheta)}}_1 \Ket{\overline{\Lambda_B'(\varphi)}}_2, \nonumber \\ \label{Eqn:EndOfProof}
\end{eqnarray}
where we use primed ket vectors on the right-hand side of the equation to capture the remaining phase $r(r-1) / 2 - |\mathbf{s}| |\mathbf{t}| = |\mathbf{s}|(|\mathbf{s}|-1) / 2 + |\mathbf{t}|(|\mathbf{t}|-1) / 2$ from Eq.~(\ref{Eqn:Expansion}).

The phase $(-1)^{AB}$ in the result shown in the above equality generates the entanglement we require to perform the teleportation as shown in Eq.~(\ref{Eqn:EntangledLogicalState}). We finally check the effect of $U$ on the two partitions of the system to verify our global unitary is suitable for our syndrome readout protocol. In particular we have that
\begin{equation}
\Ket{\overline{\Lambda_A'(\vartheta)}}_\alpha = (-\text{i}) ^A\Ket{\overline{\Lambda_A(\vartheta+\pi / 2)}}_\alpha.
\end{equation}
The above expressions follow from the fact that the new terms that appear in the coefficients of the computational basis ket vectors of state $\ket{\overline{\Lambda_A(\vartheta)}}_\alpha$ are of the form
\begin{equation}
(-1)^{|\mathbf{s}| (|\mathbf{s}| - 1) / 2} = \exp(\text{i} \pi |\mathbf{s}| / 2),
\end{equation} for even values of $|\mathbf{s}|$ and 
\begin{equation}
(-1)^{|\mathbf{s}| (|\mathbf{s}| - 1) / 2} = -\text{i}\exp(\text{i} \pi |\mathbf{s}| / 2)
\end{equation} for odd values of $|\mathbf{s}|$.

With this we have that $U$ acts on the logical space of the system expressed in the logical basis like
\begin{equation}
U = W \times \left( \overline{S}^\dagger \otimes \overline{S}^\dagger \right) \times \overline{CZ}
\end{equation}
where $\overline{S} $ is the phase gate
\begin{equation}
\overline{S} = \Ket{\overline{\Lambda_0(\vartheta)}}\Bra{\overline{\Lambda_0(\vartheta)}} + \text{i} \Ket{\overline{\Lambda_1(\vartheta)}}\Bra{\overline{\Lambda_1(\vartheta)}} 
\end{equation}
and $\overline{CZ} $ is the controlled-phase gate such that
\begin{eqnarray}
\overline{CZ} &= \Ket{\overline{\Lambda_{00}(\vartheta,\varphi)}} \!\! \Bra{\overline{\Lambda_{00}(\vartheta,\varphi)}} + \Ket{\overline{\Lambda_{01}(\vartheta,\varphi)}}\!\! \Bra{\overline{\Lambda_{01}(\vartheta,\varphi)}}  \nonumber \\ 
& \!\! + \! \Ket{\overline{\Lambda_{10}(\vartheta,\varphi)}} \!\! \Bra{\overline{\Lambda_{10}(\vartheta,\varphi)}}  -\Ket{\overline{\Lambda_{11}(\vartheta,\varphi)}} \!\! \Bra{\overline{\Lambda_{11}(\vartheta,\varphi)}}, \nonumber \\
\label{Eqn:CZ}
\end{eqnarray}
where $\ket{\overline{\Lambda_{AB}(\vartheta,\varphi)}} = \ket{\overline{\Lambda_A(\vartheta)}} \ket{\overline{\Lambda_B(\varphi)}}$ as defined above. The operator $W = \prod_{j \in \mathcal{Q}} \exp ( \text{i} \pi Z_j / 2)$ rotates the angle of the local bases of the physical qubits such that $\vartheta \rightarrow \vartheta+\pi / 2$ and $\varphi \rightarrow \varphi + \pi / 2$. Like the $V$ term that appears in Eq.~(\ref{Eqn:IntermediatePhase}), as discussed in Appendix~\ref{app:a}, this $W$ term is inconsequential provided we track the change in local basis the operator introduces.

It is important to be vigilant of the imaginary phase terms that emerge due to the $\overline{S}^\dagger \otimes \overline{S}^\dagger$ term of $U$. In particular, the states $\overline{S} \Ket{\overline{\Lambda_\pm(\vartheta)}}$ and $\overline{S}^\dagger \Ket{\overline{\Lambda_\pm(\vartheta)}}$ are not product states, and as such we cannot measure the state of this subsystem to complete the teleportation using the single-qubit measurements that are available with our proposed experimental apparatus. However, this is not problematic in practice as each partition of the system is acted upon by the global unitary exactly twice from its initialisation i.e. in the encoding and subsequent decoding steps, thus cancelling the problematic imaginary phase and we thus restore the subsystem to a basis that we can measure through single-qubit measurements. Indeed, suppose we begin in the state $\ket{C'} =  \left( S^\dagger \otimes \openone \right) \ket{\overline{\Lambda_\psi(\vartheta)}}_1 \ket{\overline{\Lambda_+(\varphi)}}_2 $ where the first subsystem supports the encoded information up to a phase gate, and the second subset has been initialised in the product state $\ket{\overline{\Lambda_+(\varphi)}}$. We find that
\begin{equation} 
 U \ket{C'}  = W \left(\left( \overline{S}^\dagger  \right)^2 \otimes \overline{S}^\dagger \right) \overline{CZ}\ket{C},
\end{equation}
where $\overline{CZ} \ket{C}$ is shown in Eq.~(\ref{Eqn:EntangledLogicalState}). Using that the states $ \left( \overline{S}^\dagger\right)^2 \ket{\overline{\Lambda_\pm(\vartheta)}} = \ket{\overline{\Lambda_\mp(\vartheta)}}$ are product states we find that we can teleport the encoded information successfully using single-qubit measurements. Repeating the syndrome readout protocol again which involves a second application of $U$ will similiarly introduce a second $\overline{S}^\dagger$ operation on the second subsystem which enables us to read this state with a measurement in a product basis. We thus see that we can carry out the desired syndrome readout via teleportation measurement as specified in our procedure. It is in this point we also see that the logical state we encode into the system at step~\ref{list:Initialise} in fact differs by a $\overline{S}^\dagger$ rotation as the the logical qubit that is initialised on the communication qubit at the beginning of the protocol is only acted upon by the global operation once. This is easily accounted for since this rotation is made only when the logical information is written on a single physical qubit before it is encoded to a larger subsystem of the crystal.

\section{Practical Considerations}
\label{app:d}

In Sec.~\ref{sec:setup} we give a general overview of the experimental configuration and ion trap layout, and in Sec.~\ref{sec:noise} we give an analysis on the capability of our system to tolerate noise in the environment and gate operations. We now discuss the likely sources of such noise by considering practical aspects of the experiment, and detail the actual operations used in performing each step of the protocol.

\subsection{Qubit Properties and Noise Bias}
\label{app:QubitSpecs}

The qubits of the system we propose are encoded in the ground-state Zeeman or hyperfine sublevels of suitable ions, often singly-charged Group-II species. The spontaneous emission lifetime of such levels is effectively infinite, but in practice is limited to several thousand seconds by background gas collisions~\cite{wang17}. The coherence time depends on the chosen qubit levels and magnetic field noise at the ion. For the magnetic-field sensitive qubits that must be used when implementing phase gates~\cite{lee05}, this is typically $\ll\SI{100}{\milli\second}$~\cite{sepiol19, goodwin16, Tan15, ruster16}, leading to typical noise bias $10^4<\eta < 10^7$.

\subsection{State Preparation and Measurement}
\label{app:LocalOperations}

We next consider the technical aspects of single-qubit measurement and preparation and the logical errors that may be introduced, as well as details of the collective local rotations necessary to perform the protocol.

\subsubsection{Z-basis state preparation and detection}
\label{app:spam}
The only local operations required by the protocols we have described are the read-out and state preparation of partitions of the qubit register. While a variety of readout techniques are used for different ion species and magnetic field strengths~\cite{wineland98, myerson08, nagerl99}, all ultimately depend on applying light resonant with a strong, dipole-allowed transition. With a suitably focused laser, this will allow us to prepare or measure qubits within a certain partition of the Coulomb crystal in the Pauli $Z$ basis.

State preparation can be achieved via optical pumping with frequency or polarisation selectivity, and can require as little as 10 scattered photons to prepare with high fidelity. The process can thus be considered near-instantaneous in the context of our protocols, taking only a fraction of a microsecond.

State detection is slower, and often more difficult to achieve with similar fidelity, limited by the ability to distinguish bright and dark qubit states. With a background of $\ll 1$ count per readout period, a $10^{-4}$ error rate for a single ion can be achieved by detecting an average of $\approx15$ photons. Assuming a typical detection efficiency of 3\% and scattering rate during readout of \SI{1E7}{\per\second}, this will be achieved with \SI{50}{\micro\second} of measurement time. For our protocol we must detect the state of multiple ions, which can be more challenging. However, if a fast scientific camera and suitable optics are available, these can be used to resolve the fluorescence from each site with low crosstalk, and the state of every qubit in the register can be detected in a measurement similar to that for the single ion described above.

We note that in more scalable approaches to ion trap system design, it may not be practical to integrate a camera and optics of suitably high quality into each trap. Furthermore, in multizone traps it can be hard to achieve the necessary field of view to resolve individual ions within chains across multiple trapping zones separated by many hundreds of microns. In this case, readout must be achieved via poisson discrimination of the overall fluorescence signal from all of the ions in a single zone. Achieving determination of the precise number of bright or dark ions in a large register requires very high statistics, due to the increasing overlap of the corresponding poisson distributions, and the time required to collect sufficient photons could limit the achievable teleportation rate. However, we note that determination of the exact number is not necessary for the successful implementation of our protocol, but merely the determination of the ‘majority vote’ outcome. Indeed, whatever the method used, it is important to note that very high single qubit readout fidelities are not necessary as discrimination errors during readout are equivalent to physical $Z$ errors, which can be identified and corrected by the code, provided the total number of errors remains sufficiently low. This relaxation makes it significantly easier to achieve a readout operation of sufficiently high fidelity, in the case that individual-ion readout is not available.

\subsubsection{Readout-induced logical errors}

During readout it is essential that the partitions which are not being measured are left unperturbed, as any scattered photons will destroy the coherence of the information they store. Depending on the read-out method used and required fidelity, hundreds to thousands of photons must be scattered from each measured ion without any scattering occurring from ions in the subset we wish to leave encoded. The lasers used for read-out must therefore be cleanly focused, and any stray scattered light minimised, but this is not too onerous a requirement given the significant distance between trap zones.

Finally, we note that the use of different ion species for the two partitions would permit perfect, independent readout and state preparation without individual addressing or shuttling, and furthermore present the possibility of continuous sympathetic cooling between teleportation operations. Several methods for multi-species entanglement have been demonstrated in recent years~\cite{Ballance15, Tan15, Hughes20}, and while these demonstrations have yet to be extended beyond two-qubit gates, there is no fundamental barrier to doing so.

\subsubsection{Microwave control}
\label{app:micro}

Following the discussion above, we assume for the purposes of this work that combined state preparation and measurement (SPAM) in the $Z$--basis can be achieved with $10^{-3}$ error per qubit. However, most of the operations used in our protocol require measurement and preparation in, e.g. the Pauli $X$ basis, achieved by applying a collective local rotation using microwaves resonant with the qubit transition frequency. We will now highlight several important but more subtle aspects of our application of microwave control in the context of our protocol.

While the protocol itself works for partitions prepared in \emph{any} equatorial bases (i.e. abitrary and different azimuthal angles $\vartheta, \varphi$ for the two partitions $\mathcal{Q}_1$ and $\mathcal{Q}_2$), if we want to perform all local rotations with a global microwave field we require $|\vartheta-\varphi|=\pi/2$, in order to allow independent rotations of the of the partitions. After the global entanglement operation, resulting in the state described in Eq.~\eqref{Eqn:EntangledLogicalState}, the two partitions will then be encoded in different, orthogonal bases, such as $X$ and $Y$. We measure one partition in a certain basis, say $X$, as follows: we rotate the entire qubit register around the $Y$ axis and measure the desired partition in the $Z$ basis via state-dependent fluorescence detection. The qubits in the other partition remain oriented in the $Y$ basis, while acquiring a relative phase: the information therefore remains protected throughout the measurement. 

The same procedure is used in Sec. ~\ref{sec:protocol}  of the main text when initialising one partition in a given basis: one partition is initialised to the Z basis, then the entire register is rotated around the axis in which the second partition is already encoded. During this operation, it is important to consider the action of the microwaves on the encoded partition: this manipulation applies a unitary operation on the logical subspace. This does not present an issue, provided we keep track of the rotation on the logical qubit.

The relative phase acquired by the logical qubit depends on the number of physical qubits in the code which have experienced a dephasing error. Since we do not assume the ability to resolve the fluorescence of individual ions, we need to manage the relative phase introduced by this operation to preserve the encoded data. Our solution to this is a three-step preparation process. We first perform the prescribed global rotation. We then pump the qubits that are undergoing initialisation back into the $\ket{0}$ state, before rotating the entire register again using an equal but opposite rotation thereby nulling the error due to any dephasing noise. This initialisation protocol is not robust to dephasing errors that occur during the process, which will lead to the improper nulling of the phase. However, this is not a major limitation as the optical pumping and microwave rotations can be performed much faster than the entangling operation itself, and make up an insignificant fraction of the total storage time. It should also be noted that this issue may be circumvented entirely by replacing global microwave control with rotations induced by Raman lasers that physically address certain subsets of the register, although this introduces considerable additional experimental complexity.

Throughout Secs.~\ref{sec:setup} \&~\ref{sec:protocol} we have assumed that encoding and measurement are always in the Pauli $Y$ and $X$ basis. However, as we have shown in Appendix~\ref{app:a}, residual single qubit rotations will lead to codewords in a more general equatorial basis $\ket{\overline{\Lambda_+(\vartheta)}},\ket{\overline{\Lambda_-(\vartheta)}}$. This provides equivalent protection against dephasing errors, but it is essential that the angle $\vartheta$ is well known. This angle will define the basis for subsequent qubit state preparation and read-out and thus the phase angle of the microwaves that must be applied. Later in this Appendix (see Subsec.~\ref{subsection:NullingRotations}) we propose a method to suppress fluctuations in $\vartheta$ due to laser intensity noise.

High fidelity readout and qubit rotations will require microwaves with an intensity and polarisation that are both stable and uniform across at least one trapping zone. While this is made easier by the long wavelength of the radiation relative to the dimensions of the ion crystals, it is not always trivial to achieve in small Paul traps, where the proximity of electrode surfaces can lead to greater local variations in the microwave field. However, provided the uncertainty in the microwave Rabi frequency across all ions can be reduced to the $10^{-2}$--$10^{-3}$ level, any contribution to the effective error rate will be negligible.

\subsection{Entangling-Gate-Induced Errors}
\label{app:ODFerrors}

We finally focus on errors that may be introduced while applying the entangling gate operation, including those more generally associated with the generation of optical dipole forces (ODFs).

The displacement drive required to produce the phase gates used in this proposal relies on the application of spin-dependent ODFs. There are a variety of ways to engineer ODFs, depending upon the species of trapped ion, Zeeman splitting and presence of hyperfine structure, and detailed discussions can be found in~\cite{wineland98, leibfried03a, britton12}. Importantly, we require a highly stable and uniform ODF across the qubit array. As such it will be necessary to position the crystal within the centre of a large (and most likely elliptical) beam. In general, the force is generated by configuring a pair of crossed laser beams with difference frequency $\mu_L$, incident at angles $\pm\theta_R/2$ symmetric about the trap axis, such that the optical lattice is aligned to the ion crystal axis and the ODF drives the transverse motional modes. The relative detuning $\mu_L$ of the crossed beams will typically be set close to the frequency of one of the trap modes, while both are also detuned several $\SI{}{\tera\hertz}$ from the nearest electronic transition. Depending on the angle of the quantisation magnetic field, beam polarisation, transition type and detuning, a variety of different ODF beam configurations are possible~\cite{ozeri07, kim08, baldwin21, sawyer12, sawyer21}. As will be discussed in Section~\ref{subsection:spontaneous}, the choice of polarisation or transition can be crucial in minimising errors due to spontaneous photon scattering.

\subsubsection{Nulling of single qubit rotation errors}
\label{subsection:NullingRotations}

In Appendix~\ref{app:a} we showed that application of a suitable ODF drive led to the desired unitary evolution, up to a global phase and an unwanted collective local $Z$ rotation on every qubit, with magnitude proportional to $F^2$. Similar local $Z$ rotations are produced directly by the optical dipole force with magnitude proportional to $F$, due to time-averaged differential AC Stark shifts between the qubit states. As we describe below, these light shifts can be nulled for suitable choice of ODF beam polarisation, but alternatively we can choose to set these such that they minimise the sensitivity of the combined $Z$ rotation angle to variations in the ODF beam intensity.

For beams with random intensity noise $\epsilon_F$, a rotation around the Z axis due to the gate will occur with an angle
\begin{equation}
\Phi\propto\left[F(1+\epsilon_F)\right]^2=c_g(1+2\epsilon_F+\epsilon_F^2)
\end{equation}
with a constant $c_g$.
The corresponding rotation due to the residual AC Stark shift during the gate period is
\begin{equation}
\Phi'\propto F(1+\epsilon_F)=c_{ac}(1+\epsilon_F),
\end{equation}
for a constant $c_{ac}$.
If, by adjusting the detuning and polarisation, we set the residual shift such that $c_{ac}=-2c_g$, we will see a net rotation of
\begin{equation}
\Phi_{tot}=-c_g (1-\epsilon_F^2),
\end{equation}
nulling our sensitivity to intensity noise during the gate, to first order.\\

\subsubsection{Errors due to spontaneous emission}
\label{subsection:spontaneous}

Decoherence due to spontaneous scattering of photons is a key consideration in all optical-dipole force driven quantum logic gates (for a detailed discussion see Ref.~\cite{ozeri07}). Decoherence can occur due to both Rayleigh scattering and inelastic Raman scattering. The latter is of most pressing concern to our protocol, because the dephasing induced by Rayleigh scattering will lead to correctable $Z$ errors, while inelastic processes are equivalent to $X$ errors against which the code offers no protection. Furthermore, for alkali-like ions heavier than magnesium, the presence of low lying $d$- or $f$- orbitals provide alternative Raman decay processes from the upper $p$ states that leave the electron outside the qubit states. Because the occurence of Rayleigh (or Raman) scattering on the dominant decay channel implies a non-vanishing $p$-level population, any non-zero scattering rate will also be accompanied by decays on these weaker channels, equivalent to uncorrectable `loss' errors. 

For appropriate choices of the ODF beam polarisation, the spin-flip Raman scattering process can be asymptotically suppressed by detuning the Raman beams far beyond the $S-P$ transition resonance, ideally to a little beyond the $p$-state fine-structure splitting~\cite{ozeri07}. In beryllium or magnesium, this allows complete suppression of unprotected errors due to ODF photon scattering, providing sufficient laser intensity is available to produce the necessary ODFs so far from resonance. For strontium ions, the total rate of unprotected errors can be reduced to $\ll10^{-4}$ over the duration of a two-qubit gate, limited by the $d-$channel Raman decays associated with residual Rayleigh scattering. For $N$-qubit gates, the duration of a gate for constant ODF intensity --- and, therefore, the error rate per qubit due to photon scattering --- scales as $\sqrt{N}$ (assuming the detuning $\delta$ from the CoM mode is adjusted to ensure we complete a single loop in phase space). The probability of a single uncorrectable error across the register therefore scales as $N^{3/2}$.

\subsubsection{Gate area errors}
\label{app:c}

We finally consider how imperfections in the global entangling operation add noise to the system. We consider the encoded state that has been prepared to teleport information from one subsystem to another with the global entangling operation $U'$ as in Eq.~(\ref{Eqn:UnitaryEvolution}). However, instead of performing the ideal operation we instead apply 
\begin{equation}
U'(T+\epsilon_t) = \exp(- \text{i} H_{\text{int}}(T+\epsilon_t)),
\end{equation}
where $T$ is the ideal pulse length and $\epsilon_t$ is some small error in its length. Given that $U'(T+\epsilon_t) = U'(\epsilon_t)U'(T)$ and $U'(T)$ acting on a code state is another code state where the two subsystems we consider within the crystal are entangled, we need only consider the unitary $U'(\epsilon_t)$ acting on code state $\left| \overline{\psi} \right \rangle$. Using that $\epsilon_t$ is small we write
\begin{equation}
U'(\epsilon_t) \approx  \mathcal{N} \prod_{j < k} \left( 1 - \text{i} \epsilon_t Z_jZ_k \right),
\end{equation}
where $ \tan \epsilon_t\approx \sin \epsilon_t \approx \epsilon_t$ and $\mathcal{N}  \approx  \left( 1 - \epsilon_t^2 \right)^{N(N-1)/2}$.

Imperfections in the unitary will only introduce an even parity of errors. We will thus look at the amplitude of the terms of the state $U'(\epsilon_t) \left| \overline{\psi} \right \rangle $ where an error lies on two particular qubits which we index $q$ and $r$. We have that the state $\left| \overline{\psi} (\epsilon_t) \right \rangle  \equiv U'(\epsilon_t) \left| \overline{\psi} \right\rangle$ is such that
\begin{equation}
\left| \overline{\psi} (\epsilon_t)  \right \rangle \approx \mathcal{N}\left( 1 -  \text{i} \epsilon_t Z_q Z_r - (N-2) \epsilon_t^2 Z_qZ_r + \dots  \right)  \left| \overline{\psi} \right\rangle .
\end{equation}
The probability of an error occuring at site $q$ and $r$ is given by $P(E_{qr})\approx \left| \left \langle \overline{\psi}\right| Z_q Z_r \left| \overline{\psi} (\epsilon_t) \right \rangle \right|^2 $ and we can easily check that
\begin{equation}
 \left \langle \overline{\psi} \right| Z_q Z_r \left| \overline{\psi} (\epsilon_t) \right \rangle  \approx - \mathcal{N}\left(  \text{i} \epsilon_t + (N-2) \epsilon_t^2 + \dots \right), 
\end{equation}
where the ellipsis represents higher order terms in $\epsilon_t$. We thus come to the effective probability of error per qubit 
\begin{equation}
\begin{aligned}
P(E_q)&\approx\sum_{r\not=q} P(E_{qr})\\
&\approx \mathcal{N}^2 (N-1)\left(\epsilon_t^2 + N^2 \epsilon_t^4  \dots \right).
\end{aligned}
\end{equation}
If we take relatively small system size such that $N \ll 1 / \epsilon_t^{2}$, we find
\begin{equation}
P(E_q)\approx N\epsilon_t^2.
\end{equation}
The effective single qubit error rate increases linearly with the number of qubits, due to the global nature of the entangling gate, and the size of the system is thus also limited by the precision to which we are able to tune the pulse length. For the purposes of this work, we assume that gate area can be controlled to within $1\%$.

\subsection{Overall Teleportation Error Model}

Here we summarise the overall teleportation error model, i.e. the code-protected $Z$-channel errors which lead to a reduction in the effective code size $n\rightarrow n_\mathrm{eff}$, and a single $X$ error channel corresponding to Raman scattering processes, which dominates the unprotected teleporation error $P_T$.

The total probability of $Z$ error per qubit during each teleportation cycle is given by
\begin{equation}
p_Z^\mathrm{T}=p_Z^\mathrm{pre}+p_Z^\mathrm{mw}+p_Z^\mathrm{area}+p_Z^\mathrm{ray}+p_Z^\mathrm{det},
\end{equation}
while the probability of $X$ error per qubit per cycle is simply
\begin{equation}
p_X^\mathrm{T}=p_X^\mathrm{ram}.
\end{equation}
The contributing components and the corresponding errors are given in Table~\ref{tab:errors}.
\begin{table}[hb]
\renewcommand*{\arraystretch}{1.3}
\centering
\begin{tabular}{|c|c|c|}
\hline
Term & Description & Value \\
\hline
\multicolumn{3}{|c|}{$Z$ errors}\\
\hline
$p_Z^\mathrm{pre}$ & $\ket{0}$ State preparation & $5\times 10^{-4}$ \\
$p_Z^\mathrm{mw}$ & Single qubit microwave rotation & $10^{-4}$ \\
$p_Z^\mathrm{area}$ & Entangling gate area & $10^{-4}\times N$ \\
$p_Z^\mathrm{ray}$ & Rayleigh scattering & $\sqrt{N / 2}\times 10^{-5}$ \\
$p_Z^\mathrm{det}$ & State detection & $5\times 10^{-4}$ \\
\hline
\multicolumn{3}{|c|}{$X$ errors}\\
\hline
$p_X^\mathrm{ram}$ & Raman scattering & $\sqrt{N / 2} \times 10^{-5}$ \\
\hline
\end{tabular}
\caption{Summary of the error contributions of the model versus total qubit number $N$, for the complete teleportation process. The contribution of each channel and scaling with qubit number is discussed in the earlier sections of this Appendix. The actual values will vary considerably between species and experiments, but for purposes of this study we assume the following: State preparation and measurement (Sec.~\ref{app:spam}) is assumed to be achieved with a modest combined error of 0.1\%. Gate area errors on the microwave single qubit rotations (see Sec.~\ref{app:micro}) and entangling gates (see Sec.~\ref{app:c}) assume pulse area noise of 1\%. Scattering errors (see Sec.~\ref{subsection:spontaneous}) assume strontium ions, an optimally chosen polarisation and operation near the asymptotic limit (i.e. high ODF laser power)~\cite{ozeri07}.}
\end{table}
\label{tab:errors}

Assuming the error rates remain small, we can then find the values of $n_\mathrm{eff}$ and $P_T$ used in Eq.~(\ref{eq:TotalErrRate}):
\begin{equation}
\begin{aligned}
n_\mathrm{eff}\approx n - N p_Z^T\\
P_T\approx N p_X^T
\end{aligned}
\end{equation}
Note that for our assumed errors, {$p_Z^\mathrm{T}\leq1\%$} for $N<90$, so the impact of $Z$-type errors in the teleportation process is insignificant and $n\approx n_\mathrm{eff}$. However, the same cannot be said of the physical $X$ errors associated with teleportation, and for large noise biases, $P_T$ is likely to be the limiting factor on the performance of the protocol.

\bibstyle{plain}

%

\end{document}